\documentclass{elsart}

\usepackage{graphicx}

\usepackage{amssymb}

\newcommand{\rgn}{($\gamma$,n)}
\newcommand{\rng}{(n,$\gamma$)}
\newcommand{\rga}{($\gamma$,$\alpha$)}
\newcommand{\rag}{($\alpha$,$\gamma$)}
\newcommand{\rgp}{($\gamma$,p)}

\newcommand{\augn}{$^{197}$Au\rgn $^{196}$Au}
\newcommand{\aung}{$^{197}$Au\rng $^{198}$Au}
\newcommand{\astph}{astrophysical}
\newcommand{\ppro}{$p$-process}

\newcommand{\gpro}{$\gamma$-process}

\newcommand{\prc}{Phys.\ Rev.\ C\ }
\newcommand{\prl}{Phys.\ Rev.\ Lett.\ }

\begin{document}

\begin{frontmatter}



\title{Photoactivation at a clinical LINAC: The \augn\ reaction slightly
  above threshold}


\author[Diak]{P.\ Mohr}
\ead{Peter.Mohr@diaksha.de}
\author[Diak]{S.\ Brieger}
\author[Diak]{G.\ Witucki}
\author[StMi]{M.\ Maetz}
\address[Diak]{Strahlentherapie, Diakonie-Klinikum Schw\"abisch Hall, \\
  Diakoniestra{\ss}e 10, D--74523 Schw\"abisch Hall, Germany}
\address[StMi]{Gymnasium bei St.\ Michael, \\
  T\"ungentaler Str.\ 92, D--74523 Schw\"abisch Hall, Germany}

\begin{abstract}
The properties of a clinical LINAC are investigated for a study of
photoactivation cross sections slightly above the neutron threshold. As an
example, the photoactivation of a tiny amount of gold by the \augn\
reaction has been measured. The derived photon intensity is at least
comparable to conventional and widely used photon sources. In
combination with its extremely stable operation, a clinical LINAC
ensures that photoactivation studies can be performed for a wide
number of targets with very limited beamtime.
\end{abstract}


\end{frontmatter}

\section{Introduction}
\label{sec:intro}
The technique of photoactivation has found applications in various
fields \cite{ACT07,Mas06}. At low-energy accelerators photoactivation
experiments are restricted to isomers (see e.g.\ \cite{Ngh98,Bel01});
only few experiments have been carried out. The majority of
photoactivation experiments is performed using bremsstrahlung with
endpoint energies of about $20 - 30$\,MeV; in this case the huge cross
section of the \rgn\ reaction around the giant dipole resonance at
energies of about 15\,MeV leads to excellent sensitivity of the
photoactivation technique \cite{Mas06}. It has been shown that medical
accelerators are able to provide a sufficient photon intensity for
such experiments \cite{Web05}. In the present investigation we focus
on photoactivation experiments at energies slightly above the neutron
threshold which are feasible at a clinical linear accelerator
(cLINAC) with relatively low photon energies up to 10\,MeV.

The experimental study of photon-induced cross sections has found
increasing interest in the last years. Most of the recent studies were
motivated by the \astph\ relevance of photon-induced reactions in the
so-called \astph\ \ppro\ or \gpro\ where neutron-deficient heavy
nuclei are synthesized by a series of \rgn , \rga , and \rgp\
reactions at temperatures of about two to three billions Kelvin ($T_9
= 2 - 3$) \cite{Woo78,Lam92,Rau02,Arn03,Hay04,Uts06}. At such high
temperatures the high-energy tail of the blackbody radiation spectrum
has a sufficient intensity to induce a noticeable reaction rate for
photon-induced reactions.

It has been shown that the \astph ly relevant energy window for \rgn\
reactions is located slightly above the threshold of the \rgn\
reaction and has a width of less than 1\,MeV \cite{MohrPLB}; the
position of this window is almost independent of the temperature
$T$. In contrast the energy window for \rga\ reactions depends
sensitively on the temperature $T$ and has a much broader width of
several MeV. The position and width can be estimated from the Gamow
window of the inverse \rag\ capture reaction. The position is shifted
by the $Q$-value of the \rag\ reaction
($E_{(\gamma,\alpha)}^{\rm{eff}} = E_{(\alpha,\gamma)}^{\rm{eff}} +
Q_{\alpha}$), and the width is practically identical as in the \rag\
reaction \cite{MohrEPJA}.

Recent experiments have been performed with almost monochromatic
photons from laser Compton scattering (LCS)
\cite{Uts06,Uts03,Shi05,Gok06,Uts06a} as well as 
with continuous photon energy distributions from bremsstrahlung
\cite{Uts06,MohrPLB,Vogt02,Sonn04,Mul06,Wag05}. In addition, a new
photon source 
using a superconducting wiggler at an electron storage ring with the
energy of several GeV has been suggested \cite{UtsNIMA,MohrEPJA}.

It is the aim of the present work to compare the performance of
previously used bremsstrahlung photon sources (e.g., S-DALINAC at TU
Darmstadt or ELBE at Forschungszentrum Dresden in Rossendorf or
commercially available electron accelerators) with the output of a
cLINAC. In the present investigation we use the cLINAC of
Diakoniekrankenhaus Schw\"abisch Hall which is a standard cLINAC
manufactured by Elekta, Crawley, United Kingdom. The measured yield
from photoactivation of a gold sample via the \augn\ reaction allows
the determination of the number of photons in the clinically used
photon beam because the cross section of this reaction has been
studied carefully in a wide energy range \cite{Vogt02,Uts03,Ber87}. We
derive the number of photons at an energy of 9\,MeV using a simplistic
linear spectral distribution and a realistic photon spectrum from a
GEANT simulation.

The paper is organized as follows. In Sect.~\ref{sec:clinac} the
general properties of the cLINAC are summarized. In
Sect.~\ref{sec:exp} the experimental set-up for the photoactivation of
gold by the \augn\ reaction is presented. In Sect.~\ref{sec:analysis}
the experimental data are analyzed, and the photon density is derived
from the measured photoactivation yield. The experimental results are
discussed in Sect.~\ref{sec:disc} and compared to other photon sources
in Sect.~\ref{sec:comp}. Finally, conclusions are given in
Sect.~\ref{sec:conc}.

\section{Properties of a clinical LINAC}
\label{sec:clinac}
The cLINAC under study is an up-to-date cLINAC which has been
installed in 2005 by the manufacturer Elekta. The following schematic
view of the cLINAC is summarized from the technical documentation of
the cLINAC \cite{Elekta}. The very stable and reliable operation of a
cLINAC under clinical conditions is an excellent basis for basic
physics research. Typically a cLINAC is running every working day. The
required maintenance is 7 days per year of operation. The manufacturer
guarantees a 96\,\% availability which corresponds to 10 further days
of unplanned maintenance per year. The overall availability of a
cLINAC is more than 90\,\%. Such a high availability has indeed been
achieved for the cLINAC under investigation in its first two years of
operation including all its peripheral devices. The availability of
the beam from the cLINAC is even better.

The primary electron beam is generated in a gun with an energy of about
50\,keV leading to an injection velocity of about 40\,\% of the speed
of light: $v \approx 0.4\,c$. The electrons are accelerated in a
copper cavity by 3\,GHz radiofrequency (RF) with a peak power of about
5\,MW in the pulse and with a pulse repetition frequency (PRF) of
200\,Hz or 400\,Hz (depending on the chosen energy). For an electron
energy of 10\,MeV a primary electron current of about 500\,mA in the
pulse can be roughly estimated assuming that 100\,\% of the RF
power are converted into electron energy. Together with the PRF of
200\,Hz and a typical pulse width of about 3\,$\mu$s this results in an
average electron current of about 300\,$\mu$A.

Steering and focusing of the electron beam is achieved by standard
magnetic and electrostatic devices. An achromatic bending system is
used to transport the beam onto a massive tungsten target where the
electron beam is stopped and bremsstrahlung with an endpoint energy
identical to the primary kinetic electron energy is produced.

\begin{figure}
\begin{center}
\includegraphics[ bb = 70 320 475 570, width = 120 mm,
clip]{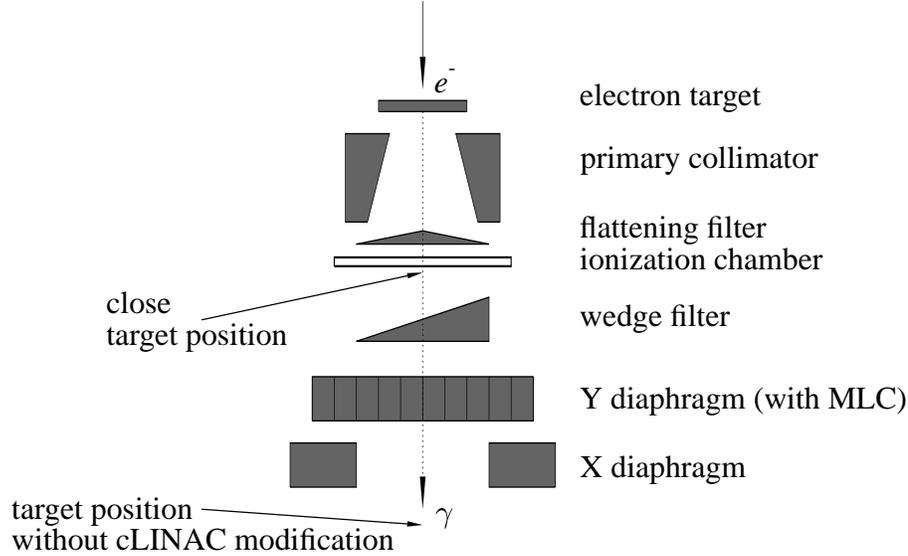}
\end{center}
\caption{
\label{fig:schematic}
Schematic view of the photon beam production and collimation of a
cLINAC. Two target positions for photoactivation are shown: a close
position at SSD $\approx 20$\,cm which requires some modifiaction of
the cLINAC and a second position at SSD $\approx 55$\,cm which can be
used in the clinical set-up (see text).
}
\end{figure}

After a massive primary collimator with a thickness of about 10\,cm
the photon beam hits a flattening filter. The bremsstrahlung emission
process is focused at forward angles. The shape of the flattening
filter is optimized to obtain an almost flat photon angular
distribution which finally has to deliver a flat dose profile to the
patient. The distance from the almost point-like photon source to an
irradiation position is called source-surface distance (SSD). A
schematic view of the bremsstrahlung production target and the
collimation system is given in Fig.~\ref{fig:schematic}.

Behind the flattening filter a thin segmented ionization chamber is
mounted. The various segments of the chamber are used to control the
position and flatness of the beam. In addition, because of the strong
sensitivity of the photon angular distribution on the electron energy,
the ratio between outer and inner segments of the ionization chamber
is used to control the energy of the primary electron beam.

Between ionization chamber and the diaphragm system a so-called wedge
filter may be used. This wedge leads to a well-defined asymmetric dose
profile.

For the shaping of the photon beam according to the requirements of
patient treatment a system of diaphragms in $X$ and $Y$ direction
allows to define rectangular fields with a maximum size of $40 \times
40$\,cm$^2$ at the nominal distance SSD = 100\,cm. Additionally, a
so-called multi-leaf collimator (MLC) allows to define irregular field
shapes according to the clinical requirements. Together with the
so-called wedge filter the diaphragm system requires about 35\,cm
space below the ionization chamber.

The ionization chamber is an essential part of the cLINAC and its beam
control system. This restricts the positioning of the target samples
for photoactivation experiments to distances larger than about 20\,cm
from the almost point-like photon source. The diameter of the primary
electron beam is typically less than 2\,mm. For this closest position
some parts of the cLINAC have to be removed; otherwise such a close
target position is not accessible. Without any modification of the
cLINAC, the closest target position is located at SSD $\approx 55$\,cm
from the point-like photon source.

A typical cLINAC is able to provide electron beams at a number of
fixed energies. The cLINAC under study is configured to provide
primary electrons at $E = 6$\,MeV (X6) and 10\,MeV (X10). A
variation of the pre-defined fixed energies is possible by readjusting
the operational parameters of the cLINAC. Because of the fully digital
operation of the cLINAC under investigation, the pre-defined clinical
beams can be restored within seconds by a software reload of the
original parameters.

\begin{figure}
\begin{center}
\includegraphics[ bb = 105 100 500 380, width = 120 mm,
clip]{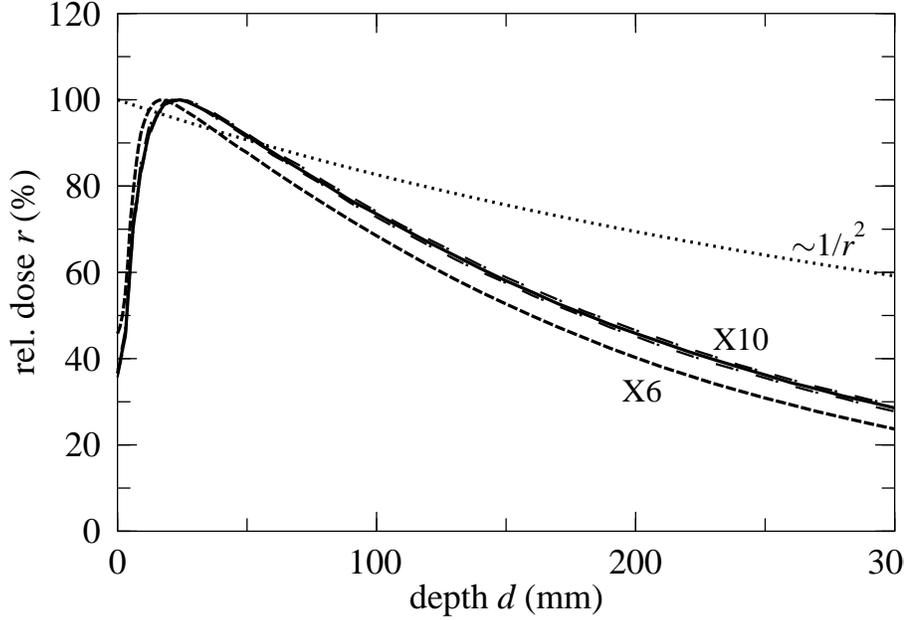}
\end{center}
\caption{
\label{fig:depth}
Measured depth dose curves for two photon energies X6 (dashed line)
and X10 (full line) corresponding to beam quality factors $Q = 0.685$
and $Q = 0.732$. The dash-dotted lines show depth dose profiles for
X10 with slightly increased $Q_> = 0.738$ and reduced $Q_< = 0.727$;
these profiles have been measured using the predefined X10 beam with
slightly modified parameters.  For comparison the depth dependence for
absorption-free radiation (proportional $1/r^2$) is also shown (dotted
line).
}
\end{figure}

The clinical energy definition of a photon beam is given by the
so-called beam quality factor which is derived from the measured depth
dose curve in a water phantom. The standard definition of the beam
quality factor $Q$ uses a set-up with SSD = 100\,cm from the
practically point-like photon source to the surface of the water
phantom and a field size of $10 \times 10$\,cm$^2$. The beam quality
factor is then defined by the ratio of doses $D_{20}$ and $D_{10}$
measured at a depth of 20\,cm and 10\,cm:
\begin{equation}
Q = D_{20} / D_{10}
\label{eq:q}
\end{equation}
Two typical depth-dose curves are shown in Fig.~\ref{fig:depth} for
the available energies X6 and X10; the corresponding beam quality
factors are $Q = 0.685$ for X6 and $Q=0.732$ for X10. It has been shown
that the stability of cLINACs is excellent. For the cLINAC under study
we find very minor variations of $Q$ of less than 1\,\% during
typical irradiations with durations of a few minutes
\cite{MohrStrOnk} and a long-time stability within $\pm 1$\,\%.

By readjustment of the X10 beam parameters we have increased and
decreased the electron energy. Two further depth dose profiles were
measured where the changes in electron energy translate to slightly
increased and reduced beam quality factors: $Q_> = 0.738$ and $Q_< =
0.727$. These changes in $Q$ correspond to a variation of the endpoint
energy of about 500\,keV. The depth dose profiles with increased $Q_>$
and decreased $Q_<$ are shown in Fig.~\ref{fig:depth} by dash-dotted
lines. After restoring the original parameters of the X10 beam, a
further depth dose curve was measured. This second curve with the
original X10 parameters typically deviates by less than 0.1\,\% from
the first X10 depth dose curve, and the beam quality factor of the
second curve is $Q = 0.732$ which is identical to the first
measurement.

Under clinical conditions the cLINAC is able to provide a dose of
slightly above 5\,Gy per minute in the dose maximum of the depth dose
curve in a water depth of about $1-2$\,cm. It is well-known that the
average photon energy of bremsstrahlung from a cLINAC is about or slightly
less than one third of the nominal endpoint energy \cite{Verh03};
i.e., for the X10 radiation in the present experiment one finds an
average photon energy of about $\overline{E} \approx 3$\,MeV. A simple
estimate of the photon intensity can be obtained as follows. From the
depth dose curve in Fig.~\ref{fig:depth} (which includes the obvious
$1/r^2$ dependence) and its comparison to the $1/r^2$ dependence it is
possible to calculate that about 2.5\,\% of the incoming photons are absorbed
in a layer of 1\,cm water. Thus, for delivery of 1\,Gy in a $10 \times
10$\,cm$^2$ layer with 1\,cm thickness a deposited energy of 0.1\,J is
required which corresponds to a primary energy of the photon beam of
$0.1\,{\rm{J}}/0.025 = 4$\,J. Together with the average photon energy
of $3\,{\rm{MeV}} = 4.8 \times 10^{-13}$\,J one finds that $8.3 \times
10^{12}$ incoming photons are required for the dose of 1\,Gy in this
layer with a volume of $V = 100$\,cm$^3$. This corresponds to $8.3
\times 10^{10}$ photons per cm$^2$ or a photon intensity of about $7
\times 10^9/({\rm{cm}}^2\,{\rm{s}})$ for the delivery of 5\,Gy in one
minute. The spectral intensity is of the order of $2 \times
10^6/({\rm{keV}}\,{\rm{cm}}^2\,{\rm{s}})$ assuming that the
main intensity of a real bremsstrahlung spectrum is located between 1
and 5\,MeV. The spectral intensity of more than
$10^6/({\rm{keV}}\,{\rm{cm}}^2\,{\rm{s}})$ is more than one order of
magnitude higher than $10^4 -
10^5/({\rm{keV}}\,{\rm{cm}}^2\,{\rm{s}})$ as e.g.\ stated in
\cite{MohrNIMA} for the S-DALINAC.

\begin{figure}
\begin{center}
\includegraphics[ bb = 100 100 510 380, width = 120 mm,
clip]{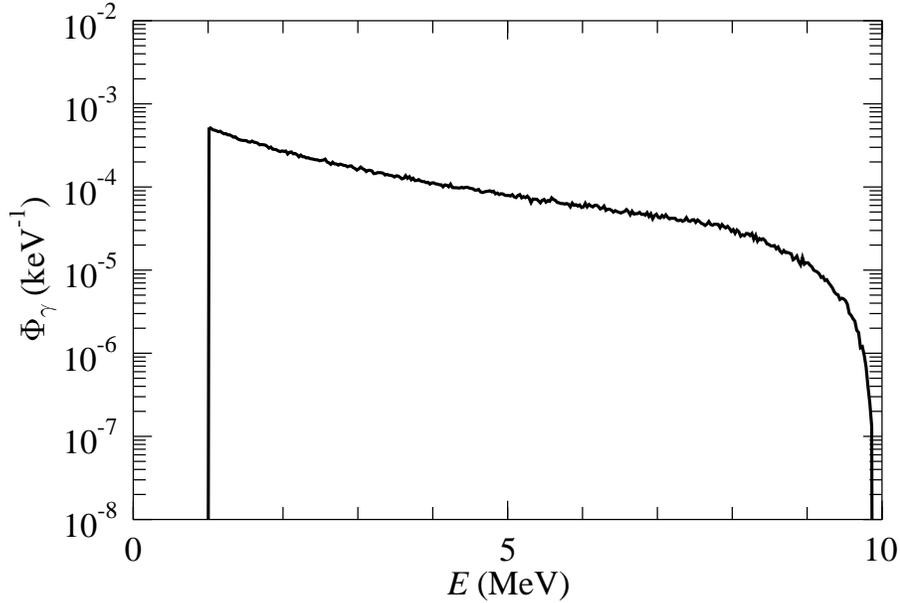}
\end{center}
\caption{
\label{fig:phigamma}
Normalized energy distribution $\Phi_\gamma(E)$ for thick-target
bremsstrahlung with an endpoint energy $E_0 = 9.9$\,MeV from a GEANT
simulation (calculation taken from \cite{Vogt02}). 
}
\end{figure}

A typical energy distribution of bremsstrahlung is shown in
Fig.~\ref{fig:phigamma}. The energy distribution has been calculated
for an electron energy of $E_0 = 9.9$\,MeV and a thick copper target
\cite{Vogt02} using the simulation code GEANT \cite{GEANT}. A lower
cutoff of $E = 1$\,MeV was used in the simulation calculation to
reduce the required time for the computation. The absorption of
photons in the electron target and the flattening filter leads to a
less sharp cutoff at similar energies in a realistic cLINAC
\cite{Verh03}. The shown energy distribution $\Phi_\gamma(E)$ is
normalized to unity:
\begin{equation}
\int_0^{E_0} \Phi_\gamma(E)\,dE = 1
\label{eq:norm}
\end{equation}
The average energy
\begin{equation}
\overline{E} = \int_0^{E_0} E\,\,\Phi_\gamma(E)\,dE
\label{eq:equer}
\end{equation}
for the shown spectrum in Fig.~\ref{fig:phigamma} is close to
3\,MeV.

There is no perfect agreement between the calculated set-up in
\cite{Vogt02} and the cLINAC of the present work. However, it has been
shown in \cite{Vogt01} that the spectral shape of the
energy distribution of bremsstrahlung depends only weakly on the
details of the set-up. A very similar shape to Fig.~\ref{fig:phigamma}
was obtained from a simulation of a cLINAC in \cite{Verh03}, and it
has been shown that the calculated spectral shape of bremsstrahlung
of cLINACs is close to measured energy spectra
\cite{Verh03,Moh85,Fad90,Fad91,Fad93,She02a,She02b}. In particular,
the spectral shape of a cLINAC close to the endpoint energy has been
analyzed in \cite{Krm02,Krm04}, and a parametrization is given that
corrects the simple Schiff formula for thin-target bremsstrahlung
\cite{Sch51} to the spectral shape of a cLINAC \cite{Krm02}.

Furthermore, a small variation in the energy of the primary electron
beam can be taken into account in a simple way by a corresponding
scaling the energy axis in Fig.~\ref{fig:phigamma} \cite{Vogt01}. Such
a scaling has been applied in the data analysis in
Sect.~\ref{sec:analysis}.

\section{Experimental Set-up}
\label{sec:exp}
Two gold samples were irradiated for the present experiment. A thin
foil with a thickness of $3.7$\,mg/cm$^2$, a size of $40 \times
40$\,mm$^2$, and a total weight of 59.2\,mg was mounted at the target
position without cLINAC modification at SSD $\approx 55$\,cm (see
Fig.~\ref{fig:schematic}). A second target consisted of a larger amount
($\approx 4$\,g) of irregularly shaped 333-gold (33.3\,\% mass
fraction of gold, other constituents are silver and copper in almost
identical mass fractions). The second target was mounted at a slightly
larger distance. The absolute photon intensity was determined from the
activity of the thin gold foil whereas the decay spectrum and the
half-life of the decaying activity were measured using the larger
target with its higher count rates.

The irradiation time is software-restricted to a maximum of about
300\,Gy corresponding to an irradiation time of about 55 minutes. This
maximum irradiation time was used to activate the two gold targets by
the \augn\ reaction using the X10 photon beam. It is possible to
extend the irradiation to longer times by a manual restart of the
beam. However, as can be seen in the following, one hour irradiation
time turned out to be sufficient for a measurable activation yield.

\begin{figure}
\begin{center}
\includegraphics[ bb = 110 100 510 380, width = 120 mm,
clip]{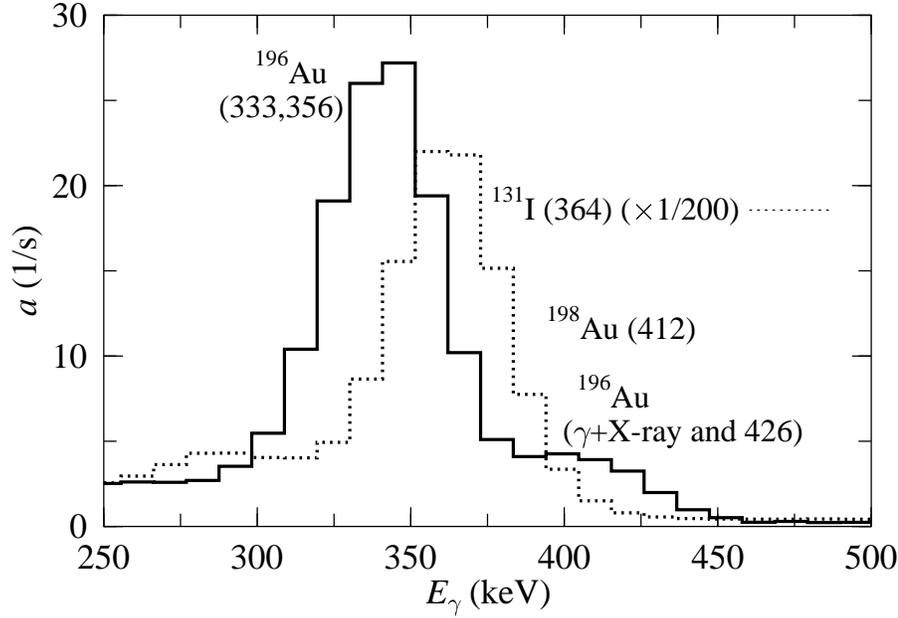}
\end{center}
\caption{
\label{fig:spek}
$\gamma$-ray spectrum of the decay of $^{196}$Au, measured with a
NaI detector. The energy resolution of the NaI detector is not
sufficient to resolve the two main lines at 333\,keV and
356\,keV. Slightly above 400\,keV a fur´ther peak can be seen which is
composed of sum lines of X-rays and 333\,keV or 356\,keV, the weak
decay line at 426\,keV, and a decay line of $^{198}$Au which was
produced by neutron capture in the \aung\ reaction. Additionally, the
calibration spectrum from $^{131}$I is shown with a dotted line.
}
\end{figure}

After irradiation the activity of the target samples was measured
with a well-shielded standard NaI(Tl) detector. The $\gamma$-ray
spectrum is shown in Fig.~\ref{fig:spek}. Although the energy
resolution of the NaI detector is not sufficient to resolve the two
main decay lines at $E_\gamma = 356$\,keV and 333\,keV, the
signal-to-background ratio is very good because the integrated
background in the shown energy range of the spectrum is below one
count per second. The properties of the NaI detector are
sufficient for the determination of the photon intensity in the
present experiment. It is obvious that the NaI detector has to be
replaced by a high-resolution detector for detailed photoactivation
studies, particularly for target nuclei with low abundances or
produced nuclei with long half-lives or weak $\gamma$-decay branches.

The absolute calibration for the NaI detector was performed with a
calibrated source of $^{131}$I. The main $\gamma$-line from the decay
of $^{131}$I at $E = 364$\,keV almost coincides with the $\gamma$-ray
energies from the decay of $^{196}$Au at $E = 356$\,keV (branching per
decay $87.0 \pm 0.8$\,\%) and 333\,keV ($22.8 \pm 0.6$\,\%); note that
the sum of the branchings exceeds 100\,\% because of cascade
transitions. For simplicity, a broad integration range of the spectrum
from about 250\,keV and 500\,keV has been used in the data
analysis. This window covers the main decay lines of $^{196}$Au and
$^{131}$I. As pointed out above, the background in this broad energy
window is below one count per second which has to be compared to the
count rate of above 100 counts per second for the larger gold sample
and more than 10\,000 counts per second for the $^{131}$I calibration
source where the deadtime of the NaI detector and its electronics is
still below 5\,\%.

\section{Data analysis}
\label{sec:analysis}
\begin{figure}
\begin{center}
\includegraphics[ bb = 105 100 510 380, width = 120 mm,
clip]{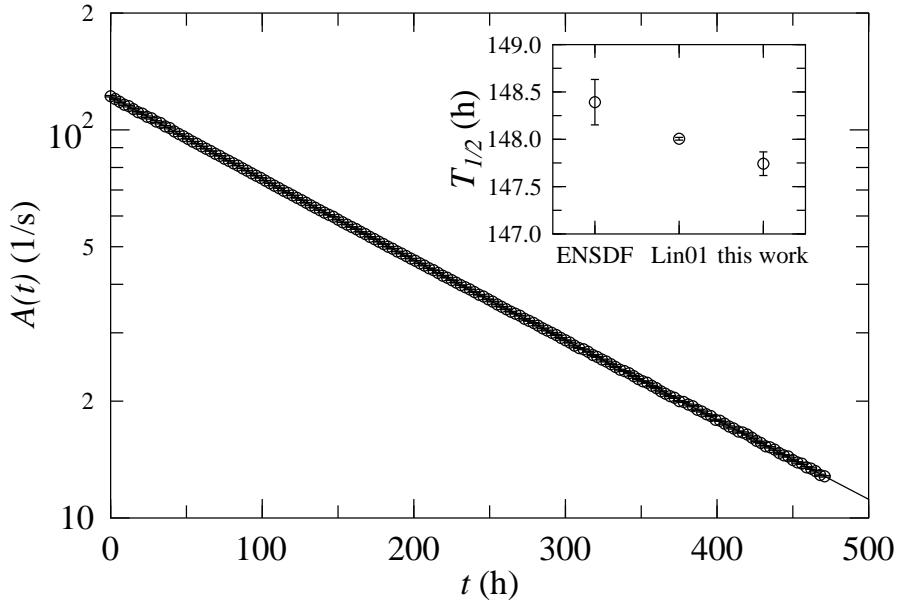}
\end{center}
\caption{
\label{fig:half}
Decay curve of the activity of $^{196}$Au measured over more than
three half-lives. The inset compares our new result to the adopted
value from the ENSDF data base \cite{ENSDF,Ike63} and the recently
measured precise result of \cite{Lin01}.
}
\end{figure}

In a first part of the data analysis the activity of the gold sample
has been measured for more than three half-lives. The result is shown
in Fig.~\ref{fig:half}. The obtained half-life of $^{196}$Au is
$T_{1/2} = 147.75 \pm 0.13$\,h which is in rough agreement with the
adopted value of $148.39 \pm 0.24$\,h \cite{ENSDF,Ike63} and overlaps
within $2\,\sigma$ with the recently measured precise number of
$148.006 \pm 0.015$\,h \cite{Lin01}. It is interesting to note that a
small component ($< 5$\,\%) with a shorter half-life around 3 days is
required for an excellent fit ($\chi_{\rm{red}}^2 \approx 1$) of the
decay curve; this component comes probably from the decay of
$^{198}$Au which can be produced by the \aung\ reaction. A very weak
decay line of $^{198}$Au was also observed in the $\gamma$-ray
spectrum of \cite{Vogt02}. The neutrons may have been produced in the
target sample itself by the \augn\ reaction and/or in the tungsten
diaphragms of the cLINAC by the $^{183,186}$W\rgn $^{182,185}$W
reactions.

The calculation of the photon intensity was performed as follows. The
activity of the thin gold foil was measured after the irradiation to
be $3.71$ counts per second. The measured activity was converted to
the number of produced $^{196}$Au nuclei taking into account the decay
constant of $^{196}$Au, the $\gamma$-ray branching in the decay of
$^{196}$Au and $^{131}$I which was used for the detector calibration,
and the detector efficiency. The total number of \augn\ reactions and
produced $^{196}$Au nuclei was $(2.38 \pm 0.25) \times 10^7$ where the
uncertainty is dominated by the absolute calibration of the NaI
detector. 

The cross section of the \augn\ reaction is well-known from the
threshold at 8.071\,MeV up to the giant dipole resonance. We use the
parametri\-zation of \cite{Vogt02} which is based on the
bremsstrahlung and photoactivation experiment of \cite{Vogt02}, one
recent data point using LCS photons and direct neutron detection
\cite{Uts03}, and positron-in-flight annihilation experiment by
\cite{Ber87}. The uncertainty of the \augn\ reaction cross section is
smaller than 5\,\% over the whole energy range of the present experiment.

\begin{figure}
\begin{center}
\includegraphics[ bb = 90 100 550 380, width = 120 mm,
clip]{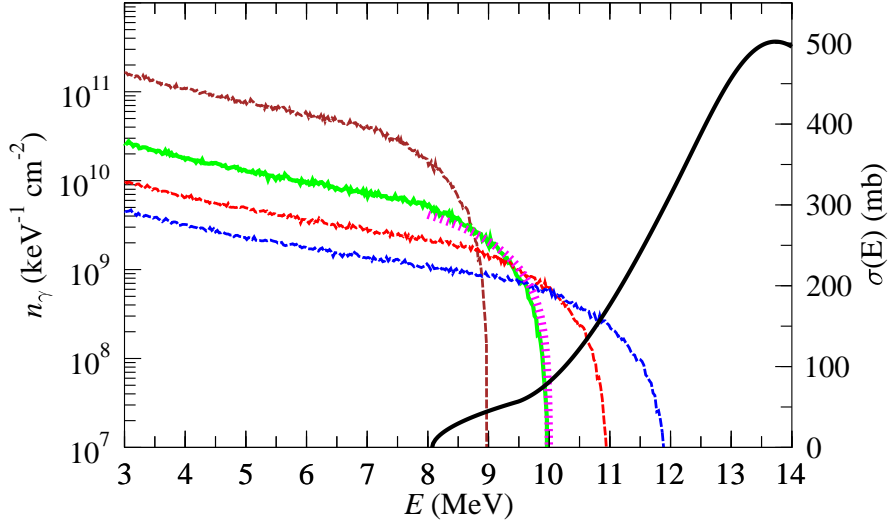}
\end{center}
\caption{
\label{fig:yield}
Number of photons $n_\gamma(E,E_0)$ at SSD $\approx 55$\,cm determined
from the photoactivation yield in the \augn\ reaction. The thick lines
show the cross section of the \augn\ reaction in the parametrization
of \cite{Vogt02} (right scale) and the photon intensity
$n_\gamma(E,E_0)$ for an endpoint energy of $E_0 = 10$\,MeV (left
scale). Additionally, the dashed lines show the results for lower
($E_0 = 9$\,MeV) and higher ($E_0 = 11$ and 12\,MeV) endpoint energies
for comparison. The hatched line represents a simplistic linear
spectral shape between 8 and 10\,MeV.
}
\end{figure}

The cross section, photon intensity, and number of reactions are
related by the yield integral
\begin{equation}
N_R = k \, N_T \times \int_{S_{\rm{n}}}^{E_0} \sigma(E) \,
\Phi_\gamma(E,E_0) \, dE 
\label{eq:yield}
\end{equation}
with the number of reactions $N_R$, number of target atoms in the
sample $N_T$, normalized bremsstrahlung spectrum $\Phi_\gamma(E,E_0)$
at endpoint energy $E_0$, cross section $\sigma(E)$, and a scaling
factor $k$ for the normalized bremsstrahlung spectrum. The integral in
Eq.~(\ref{eq:yield}) is limited by the neutron separation energy
$S_{\rm{n}}$ and the endpoint energy $E_0$. The number of photons
$n_\gamma(E,E_0)$ (per keV and cm$^{-2}$) for this irradiation is
finally obtained by
\begin{equation}
n_\gamma(E,E_0) = k \times \Phi_\gamma(E,E_0)
\label{eq:k}
\end{equation}
The scaling factor $k$ has been adjusted to reproduce the experimental
yield. The result is shown in Fig.~\ref{fig:yield} for the
distribution $\Phi_\gamma(E,E_0 = 10\,{\rm{MeV}})$ at SSD $\approx
55$\,cm (see also Fig.~\ref{fig:phigamma}). The beam quality factor of
$Q = 0.732$ which was measured in the depth dose curve (see
Fig.~\ref{fig:depth}) indicates that the endpoint energy must be close
to $E_0 = 10$\,MeV.

For experiments slightly above the \rgn\
threshold the results depend sensitively on the endpoint energy $E_0$
in the experiment. Therefore, the above procedure for the determination
of the scaling factor $k$ and the number of photons $n_\gamma(E,E_0)$
has been repeated using four normalized bremsstrahlung spectra
$\Phi(E,E_0)$ with different endpoint energies $E_0 = 9$, 10, 11, and
12\,MeV which were obtained from the basic $\Phi_\gamma(E,E_0 =
9.9\,{\rm{MeV}})$ at the endpoint energy $E_0 = 9.9$\,MeV by a simple
scaling of the energy axis (see Fig.~\ref{fig:phigamma}). The results
are also shown in Fig.~\ref{fig:yield} (dotted lines). It is obvious
that a lower (higher) number of photons at higher (lower) endpoint
energies $E_0$ is required to fit the experimental photoactivation
yield. 

As a further check we used also a simplistic linear spectral shape
between the energies of 8 and 10\,MeV. The derived photon intensity
from this simplistic spectral shape agrees at energies around 9\,MeV
within less than 10\,\% with the result using the spectral shape from
the GEANT simulation with endpoint energy $E_0 = 10$\,MeV.

Using the realistic bremsstrahlung spectrum at $E_0 = 10$\,MeV, we
find a number of photons of about $n_\gamma \approx 2 \times
10^{9}$/(keV\,cm$^2$) at energies around 9\,MeV for SSD $\approx
55$\,cm. These photons were irradiated within less than one hour. This
corresponds to a photon intensity of about $6 \times
10^5$/(keV\,cm$^2$\,s) at $E = 9$\,MeV and SSD $\approx
55$\,cm. Because of the spectral shape of the bremsstrahlung spectrum,
the number of photons at lower energies is much higher. E.g., around
$E = 3$\,MeV we find $n_\gamma \approx 3 \times 10^{10}$/(keV\,cm$^2$)
at SSD $\approx 55$\,cm which corresponds to an intensity of slightly
below $10^{7}$/(keV\,cm$^2$\,s) at this SSD.

In Sect.~\ref{sec:clinac} a rough estimate of the photon intensity of
$2 \times 10^6/({\rm{keV}}\,{\rm{cm}}^2\,{\rm{s}})$ around 3\,MeV was
given for the distance of 100\,cm between point-like photon source and
water phantom (SSD = 100\,cm). Consequently, for the smaller SSD of
$\approx 55$\,cm of the photoactivation experiment one expects about a
factor of four higher photon intensity, i.e.\ $8 \times
10^6/({\rm{keV}}\,{\rm{cm}}^2\,{\rm{s}})$. This rough estimate from
the dose absorption in a water tank is in excellent agreement with the
number of slighly below $10^{7}$/(keV\,cm$^2$\,s) which was derived in
the previous paragraph using the brems\-strahlung spectrum with $E_0 =
10$\,MeV. If one uses instead the bremsstrahlung spectrum at $E_0 =
9$\,MeV (11\,MeV), one finds a photon intensity which is a factor of
about five higher (three lower). This finding demonstrates the sensitivity
of photoactivation yields at endpoint energies slightly above the \rgn\
reaction threshold and confirms that the bremsstrahlung spectrum of
the cLINAC has an endpoint energy close to $E_0 = 10$\,MeV.

Further study of the bremsstrahlung spectrum of the cLINAC is required
to reduce the uncertainties of the above data analysis which are of
the order about 25\,\% for the derived number of photons. For future
experiments it might be helpful to establish a standard for the
normalization of experiments which consists of several nuclei with
different thresholds for the \rgn\ reaction. This can be done by
measurements with quasi-monochromatic photons. At present precise data
down to the threshold of the \rgn\ reaction are available e.g.\ for
$^{197}$Au with $S_n = 8071$\,keV \cite{Vogt02,Uts03}, $^{187}$Re with
$S_n = 7363$\,keV \cite{Shi05,Mul06,MohrPOS}, and $^{186}$W with $S_n
= 7194$\,keV \cite{Mohr04,Shi05,Sonn04}.

\section{Discussion}
\label{sec:disc}
The number of photons $n_\gamma(E,E_0)$ -- as shown in
Fig.~\ref{fig:yield} for the SSD $\approx 55$\,cm -- is sufficient to
obtain a reasonable yield in \rgn\ experiments with measuring times of
the order of one hour. Such experiments can be done easily at a cLINAC
without major efforts for preparing the experiment. Reducing the SSD
by almost a factor of three from 55\,cm to about 20\,cm at the close
target position (see Fig.~\ref{fig:schematic}) leads to an increased
photon intensity by almost one order of magnitude compared to the SSD
of this experiment. However, this reduction of the SSD requires some
preparation time and mechanical modifications which cannot be done
easily at a cLINAC in daily clinical use.

A variation of the electron beam energy requires modification and
optimization of the beam parameters which can be controlled by
simultaneous measurements of depth dose profiles. Note that the beam
quality factor of the depth dose profile is an excellent measure for
the electron beam energy (see also Fig.~\ref{fig:depth}). Although
such modifications of the electron beam are possible in principle, the
required effort and patient safety considerations will hamper larger
series of photonuclear experiments at any cLINAC in routine clinical
use because it must be absolutely sure that any modification of beam
parameters for an experiment is removed before clinical use of the
cLINAC. However, because the required beam time for photonuclear
experiments is quite small, it seems nevertheless possible to perform
larger series of photonuclear experiments e.g.\ during the set-up of a
new cLINAC at the test facility of the manufacturer or at a newly
installed cLINAC before clinical use.

Finally, radiation safety issues have to be considered before any
larger experimental programs. Typically, the radiation protection of a
cLINAC is calculated for a total dose of 1000\,Gy per week (delivered
at SSD = 100\,cm) which is by far sufficient to treat about 50
patients per day with doses of 2\,Gy. In clinical use, most of the
time is required for patient positioning, and the real irradiation
time is quite short (typically far less than one minute per
patient). The single irradiation of this experiment took about one
hour and required a total dose of about 300\,Gy. Such an irradiation
corresponds to about one third of maximum dose per week, and obviously
it cannot be performed several times a week.

However, the radiation safety problem may be solved or at least
dramatically reduced. In clinical operation, the cLINAC is rotated
around the patient. Radiation protection is designed so that 1000\,Gy
per week can safely be irradiated in any direction. Often one can find
directions with much lower radiation exposure to the surrounding
area. E.g., for the cLINAC under study the radiation exposure to the
operators varies by about a factor 5 depending on the irradiation
angle. It might be discussed with the regulatory authority to extend
the radiation time by this factor of 5 at a well-defined radiation
direction with low exposure to the operators. Additionally, the access
to areas with high exposures may be prohibited or restricted in time.

\section{Comparison to other electron accelerators and photon sources}
\label{sec:comp}
The photon intensity which can be obtained without any modifications
at SSD $\approx 55$\,cm is about one order of magnitude higher than
the available intensity at the regular irradiation position at the
S-DALINAC accelerator in Darmstadt \cite{MohrNIMA}. Here simultaneous
photon scattering \cite{Kne96} can be measured to reduce the
uncertainties of the spectral shape of the photon spectrum. At a
second irradiation position close to the photon source at S-DALINAC
one finds a photon intensity which is a factor of 300 higher than at
the standard irradiation position. This higher intensity is comparable
to the increased intensity at a smaller SSD of the cLINAC of the
present investigation.

The ELBE accelerator of Forschungszentrum Dresden is able to provide
huge electron beam currents. However, at the standard irradiation
position for nuclear resonance fluorescence experiments a thin
radiator target is used leading to photon intensities of the order of
$5 \times 10^4$/(keV\,s) at $E = 7$\,MeV for bremsstrahlung with $E_0
= 10$\,MeV \cite{FZDWeb}. A much higher photon intensity can be
obtained at the beam dump of the high-current electron beam. A photon
flux of ``up to $10^{10}$\,\,cm$^{-2}$\,s$^{-1}$\,MeV$^{-1}$'' is
stated in \cite{Erh06} but the corresponding photon energy is not
given in \cite{Erh06}. The determination of the energy spectrum of the
bremsstrahlung and the 
absolute normalization of the bremsstrahlung are more complicated
compared to the standard irradiation position. Nevertheless, a few
remarkable experiments have been performed at ELBE. In particular, first
experimental data for \rga\ reactions have been obtained \cite{Wag05}.

There are a number of commercially available electron accelerators
which may be equipped with a massive target for bremsstrahlung
production. Electrostatic accelerators like dynamitrons and
van-de-Graaf or pelletron accelerators provide high electron currents;
however, the available energy is often limited. In particular, this is
the case for single-ended electron accelerators. E.g.\ the S-Series
pelletron from National Electrostatics Corporation (NEC) (``S'' for
single-ended) is limited to electron energies below 5\,MeV
\cite{NEC}. The same limit of 5\,MeV holds for the Radiation Dynamics
(RDI) dynamitron \cite{RDI}. Because of the limited electron energy
photoactivation is practically restricted to isomers
\cite{Ngh98,Bel01}.

A high-power, high-energy electron accelerator, the so-called
rhodotron accelerator (TT series), is available from Ion Beam
Applications (IBA) \cite{IBA} with energies of 10\,MeV and beam power
up to almost 200\,kW. However, the only commercially available
rhodotron with variable energy is limited to 7.5\,MeV which is again
too low for photoactivation experiments. Some development is needed
for the 10\,MeV rhodotron as an excellent energy-variable photon
source. An energy-variable electron beam has to be implemented which
is a complicated task, in particular for continuously varying electron
energies.

Any comparison with quasi-monochromatic photon sources as e.g.\ AIST
\cite{Ohg91} at Tsukuba or HI$\gamma$S \cite{Wel03} at Duke
University, Durham, is difficult because the spectral intensity of
quasi-monochromatic photon soures is very high (or infinity for an
ideal monochromatic photon source). However, the absolute number of
photons is typically lower compared to bremsstrahlung facilities. It
depends on the requirements of the respective experiment whether a
quasi-monochromatic photon source or a bremsstrahlung photon source is
to be preferred.

Recently, a new high-intensity non-monochromatic photon source was
suggested using synchrotron radiation from a 10\,T superconducting
wiggler at a many-GeV electron storage ring at SPring-8
\cite{UtsNIMA}. A first proposal has been accepted, and hopefully
this first part of the MeV photon source will become
operational within the next years. The properties of this
outstanding photon source cannot be reached by a cLINAC but it can be
expected that the beamtime at such a huge facility will be very
limited.

\section{Conclusions}
\label{sec:conc}
A photon intensity of about $6 \times 10^5$/(keV\,cm$^2$\,s) at $E =
9$\,MeV can be achieved at the cLINAC under investigation without any
modification to the clinical set-up. The photon intensity may be
increased by almost one order of magnitude by minor
modifications. This intensity is at least comparable to photon sources
which have been used in recent photonuclear experiments.

Furthermore, the excellent stability and reproducibility of the beam
of a cLINAC are good prerequisites for such experiments. The required
beam times are short and can be done during the night when most
cLINACs are not in operation.

In summary, a cLINAC is an excellent photon source for photonuclear
experiments. Besides the radiation safety problem, there is not any
physical argument that stands against photonuclear research at a
cLINAC.

\section*{Acknowledgments}
\label{sec:ack}
We thank Th.\ Aller for his assistance and patience during the
measurements of the $\gamma$-ray spectra, the group {\it{Physik in
der Medizin}} from Gymnasium St.\ Michael (M.\ Amend, A.\ Angerer, J.\
Bahm\"uller, J.\ F\"orstel, B.\ Franz, N.\ Pohl, J.\ Scheu, A.\ Sebek,
L.\ Zipperer) for their help in the experiment, and Th.\ Preisendanz
for the support of this project.

\end{document}